\documentclass[runningheads]{svmult} 
\usepackage{makeidx}   
\usepackage{graphicx}  
\usepackage{subeqnar}  
\usepackage{multicol}  
\usepackage{cropmark} 
\usepackage{physprbb}  
\def\lsim{\, \lower2truept\hbox{${< 
\atop\hbox{\raise4truept\hbox{$\sim$}}}$}\,}   
 
\def\gsim{\, \lower2truept\hbox{${> 
\atop\hbox{\raise4truept\hbox{$\sim$}}}$}\,} 
 
\begin{document} 

\title*{The Distribution of Dark Matter in Galaxies: \protect\newline 
Constant--Density Dark Halos Envelop the Stellar Disks}

\titlerunning{ Constant Density Dark Halos Envelop Galaxy Disks}
 
\author{Paolo Salucci\inst{1}
\and Annamaria Borriello\inst{1}}
 
\authorrunning{Salucci  \& Borriello}

\institute{(1) International School for Advanced Studies SISSA-ISAS -- Trieste, I}

\maketitle

\begin{abstract}

In this  paper  we review  the  main and the most recent evidence for the 
presence of a core radius in  the distribution of the dark matter around 
spiral galaxies. Their rotation curves, coadded according to the galaxy luminosity,
conform  to an Universal  profile which can be represented as the sum of an
exponential thin disk term plus a spherical halo term with a flat density
core. From dwarfs to giants, these halos feature a constant density region of
size $r_0$ and core density $\rho_0$ related by $\rho_0= 4.5 \times 10^{-2} (r
_0/{\rm kpc})^{-2/3} {\rm M}_{\odot} {\rm pc}^{-3}$.
 At the highest masses
$\rho_0$ decreases exponentially,
 with $r_0$ revealing a lack of objects with
disk masses $> 10^{11}{\rm M}_\odot$ and central densities $> 1.5 \times
10^{-2}(r_0/{\rm kpc})^{-3} {\rm M}_{\odot} {\rm pc}^{-3}$ , which implies a
{\it maximum} mass of $\approx 2 \times 10^{12} {\rm M}_{\odot}$ for halos
hosting  spirals. The fine structure of dark matter halos is obtained from
the kinematics of 
 a number of suitable low--luminosity disk galaxies. The
inferred halo circular velocity  increases linearly with radius out
to the edge of the stellar disk, implying a constant dark halo density over
the entire disk region.  The structural properties of halos around normal
spirals are similar to those around dwarf and low surface brightness
galaxies; nevertheless they provide far more substantial evidence of the
discrepancy between the mass distributions predicted in the Cold Dark Matter
scenario and those actually detected around galaxies.  
\end{abstract} 

\section{Introduction}

Rotation curves (RC's) of disk galaxies are the best probe for dark matter
(DM) on galactic scale. Notwithstanding the impressive amount of knowledge
gathered in the past 20 years, only very recently we start to shed light to crucial aspects 
of the mass {\it distribution} including the actual density profile of dark
halos and its claimed universality.

On the cosmological side, high-resolution cosmological N-body simulations have
shown that cold dark matter (CDM) halos achieve a specific equilibrium
density profile [16 hereafter NFW, 6, 10, 14, 11]. This can be characterized
by one free parameter, e.g. ${\rm M}_{200}$, the halo mass contained within
the radius inside which the average over-density is 200 times the critical
density of the Universe at the formation epoch. In their innermost region the
dark matter profiles show some scatter around an average profile which is
characterized by a power-law cusp $\rho \sim r^{-\gamma} $, with $\gamma
=1-1.5$ [16, 14,  2].In detail, the DM density profile is:
   
\begin{equation}  
\rho_{\rm NFW}(r) = \frac{\rho_s}{(r/r_s)(1+r/r_s)^2} 
\end{equation} 
where $r_s$ is a characteristic inner radius and $\rho_s$ the 
corresponding density. Let us define the halo virial radius $R_{\rm vir}$ as the radius 
within which the mean density is $\Delta_{\rm vir}$ times the mean universal 
density $\rho_m$ at that redshift, and the associated virial mass $M_{\rm 
vir}$ and velocity $V_{\rm vir} \equiv G M_{\rm vir} / R_{\rm vir}$. By 
defining the concentration parameter as $c_{\rm vir} \equiv R_{\rm vir}/r_s$ 
the halo circular velocity $V_{\rm CDM}(r)$ takes the form [2]: 

\begin{equation} 
V_{\rm CDM}^2(r)= V_{\rm vir}^2 \frac{c_{\rm vir}}{A(c_{\rm vir})} 
\frac {A(x)}{x} 
\end{equation} 
where $x\equiv r/r_s$ and $A(x)\equiv \ln (1+x) - x/(1+x)$. 
As the relation between $V_{\rm vir}$ and $R_{\rm vir}$ is fully 
specified by the background cosmology, we assume the currently 
popular $\Lambda$CDM cosmological model, with $\Omega_m = 0.3$, 
$\Omega_{\Lambda} =0.7$ and $h=0.75$, in order to reduce from three to two 
($c_{\rm vir}$ and $r_s$) the independent parameters characterizing the model. 
According to this model, $\Delta_{\rm vir} \simeq 340$ at $z \simeq 0$. 
Let us stress that a high density $\Omega_m=1$ model, with a 
concentration parameter $c_{\rm vir}>12$, is definitely unable to 
account for the observed galaxy kinematics [13]. 
Until recently, due to both the limited number of suitable RC's and to
uncertainties  on the exact amount of luminous matter in the innermost regions
of spirals, it has been difficult to investigate the internal structure of
their dark halos. However, as a result of substantial observational and
theoretical  progresses, we have recently derived the main features of their
mass distribution for {\it a}) the
Universal Rotation Curve [20] built by coadding  $~ 1000$ RC's and {\it b}) a
number of suitably selected RC's [1].    

\section{The URC and CDM Halos}    

The assumed (and well supported) framework is: {\it a}) the mass in spirals
is distributed  according to the Inner Baryon Dominance (IBD) regime: there
is  a characteristic transition radius  $R_{IBD} \simeq 2  R_d (V_{opt}/220
\ {\rm km/s})^{1.2}$  ($R_d$ is the disk scale-length and $V_{opt} \equiv
V(R_{opt})$) according which, for $r\leq R_{IBD}$, the luminous matter
totally  accounts for the mass distribution, whereas, for $r> R_{IBD}$, DM
{\it rapidly}  becomes the dominant dynamical component [26, 24, 1].   Then,
although the dark halo might extend down to the galaxy   center, it is only
for $r>R_{IBD}$ that it gives a non-negligible contribution to the  circular
velocity. {\it b}) DM is  distributed in a different way with respect to any
of the various baryonic  components [20, 7],  and {\it c}) HI contribution to
the  circular velocity at $r< R_{opt}$, is negligible [e.g. 21].   
\subsection{Halo Density Profiles} 
 
\begin{figure}[h] 
\begin{center} 
\vspace{1truecm}
\includegraphics[width=.5\textwidth]{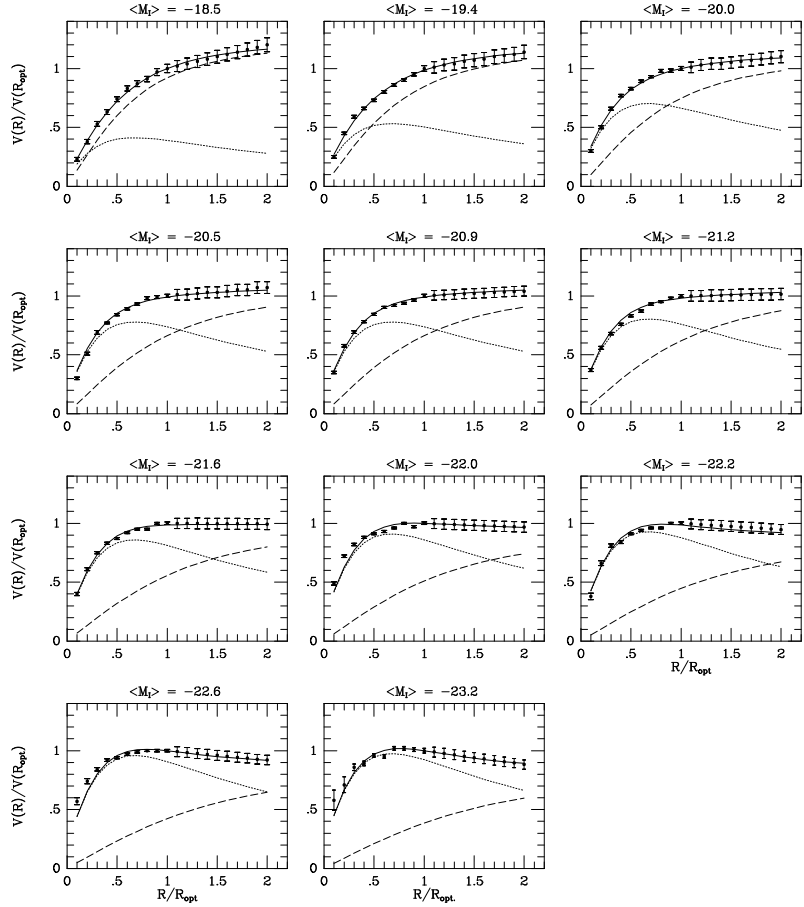} 
\end{center} 
\caption[]{Synthetic rotation curves (filled circles with error bars)
and URC (solid line) with its separate dark/luminous   contributions 
(dotted line: disk; dashed line: halo). See [20] for details} 
\end{figure} 
  
Reference [20]  have derived  from $15000$  velocity measurements of $~1000$
RC's, the synthetic rotation velocities of spirals
$V_{syn}({r\over{R_{opt}}}, {L_I\over{L_*}})$,  sorted by   luminosity (Fig.
1, with $L_I$ the  $I$--band luminosity and $L_I/L_*=10^{-(M_I+21.9)/5})$. 
Remarkably, {\it individual} RC's have   a very small variance with respect to
the corresponding synthetic curves  [20, 21, 22]:  spirals sweep a very narrow
locus in the   RC-profile/amplitude/luminosity space. On the other hand, the
galaxy kinematical  properties  significantly change with luminosity [e.g.
20], so it is   natural  to relate the mass distribution with this quantity. 
The whole set of synthetic RC's has been reproduced  by means of  the
Universal Rotation Curve (URC) $V_{URC}(r/R_{opt}, L_I/L_*)$   which includes:
{\it a}) an exponential thin disk term [9]: 

\begin{equation} 
V^2_{d,URC}(x)=1.28~\beta V^2_{opt}~ x^2~(I_0K_0-I_1K_1)|_{1.6x} 
\end{equation} 
and {\it b}) a spherical halo term: 

\begin{equation} 
V_{h,URC}^2(x)= V^2_{opt} ~(1-\beta) ~(1+a^2) {x^2 \over{(x^2+a^2})}\,, 
\end{equation} 
with $x \equiv r/R_{opt}$,  $\beta  \equiv
(V_{d,URC}(R_{opt})/V_{opt})^2$, $V_{opt} \equiv V(R_{opt})$ and   $a$ the
halo core radius in units of $R_{opt}$. At high luminosities,  the
contribution from a bulge component has also been  considered.

\begin{figure}[h] 
\begin{center} 
\vspace{-3.8truecm}
\includegraphics[width=.7\textwidth]{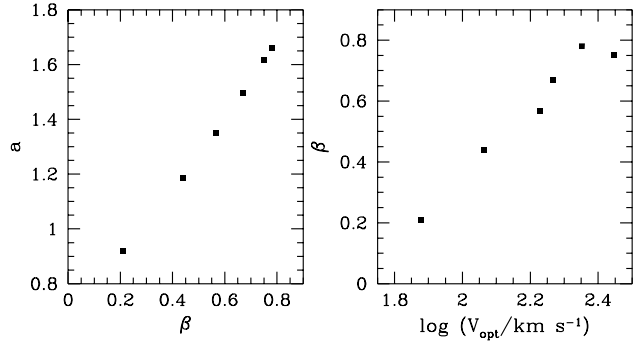} 
\end{center} 
\caption[]{$a$ {\it vs} $\beta$ and  $\beta$ {\it vs} $V_{opt}$} 
\end{figure} 
 
Let us stress that the halo velocity functional form (4) does not bias  
the mass model: it can equally  account for  maximum--disk, solid--body,
no--halo, all--halo,  CDM and core--less halo mass models. In practice, 
the synthetic curves $V_{syn}$ select the actual model out of the family of
models  $V_{URC}^2(x)=V^2_{h,URC}(x,\beta,a)+ V^2_{d, URC}(x,\beta) $, 
where  $a$ and $\beta$ are free parameters.
Adopting  $a \simeq 1.5 ( {L_I/L_*} )^{1/5}$ and  $\beta ~\simeq 0.72 +~0.44
\log ({L_I / L_*})$ [20] or, equivalently, the corresponding  $a =a(\beta)$ and
$ \beta=\beta (\log  V_{opt})$ plotted in Fig. 2,  the URC   reproduces the
synthetic curves $V_{syn}(r)$ within their r.m.s. (see Fig. 1). More in
detail, at any  luminosity and radius, $|V_{URC}-V_{syn}| <   2\%$ and the
$1\sigma$ fitting uncertainties on  $a$ and $\beta$ are about  20\% [20].    

To cope with this observational evidence  and conveniently frame the halo
density  properties, we adopted the empirical profile proposed by Burkert [3]: 

\begin{equation} 
\rho_b(r) = \frac{\rho_0 r^3_0}{(r+r_0)(r^2+r_0^2)} 
\end{equation} 
where $\rho_0$ and $r_0$ are free parameters which represent 
the central DM density and the scale radius.  Within spherical symmetry, 
the mass distribution   is given by:

\begin{equation} 
M_b(r) = 4 M_0 \{ \ln (1 + r/r_0)  -\arctan (r/r_0) + 0.5  \ln  [1
+(r/r_0)^2]\}
\end{equation}  
with $M_0$, the dark  mass within the core, given by  $M_0 = 1.6 \rho_0
r_0^3 $. The halo contribution to the circular velocity is then:

\begin{equation} 
V^2_b(r) =GM_b(r) r
\end{equation}
Although the dark matter core parameters $r_0$, $\rho_0$ and $M_0$ are 
in principle independent, the observations reveal a clear correlation 
[3]: 

\begin{equation} 
M_0 = 4.3\times 10^7 \left( \frac{r_0}{kpc} \right)^{7/3} M_{\odot}  
\end{equation} 
which, together with the above relationship, indicates that dark halos
represent  a 1--parameter family which is completely specified, e.g. by the
core mass.

\begin{figure}[h] 
\begin{center} 
\includegraphics[width=.75\textwidth, height=0.71\textwidth]{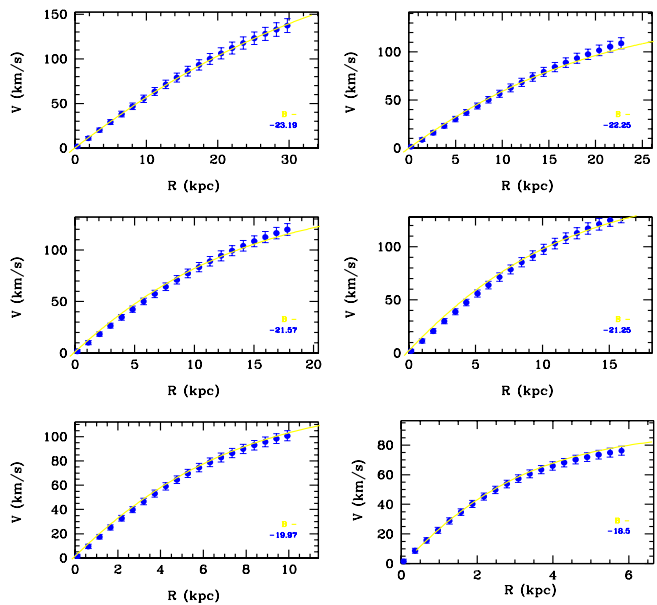} 
\end{center} 
\caption[]{URC-halo rotation curves  ({\it filled circles} with error bars)
and the Burkert model ({\it solid line}).  The bin magnitudes are also indicated} 
\end{figure}

We then compare the dark halo velocities obtained with (3) and (4), 
with the Burkert velocities $V_b(r)$ of (5)-(7),   
leaving $\rho_0$ and $r_0$ as free parameters, i.e. we do not impose  
the relationship (8). 
The results are  shown in Fig. 3: at any luminosity, out to the outermost
radii ($\sim 6 R_d$),  $V_b(r)$ is indistinguishable from $V_{h,URC}(r)$. More
specifically, by  setting $V_{h, URC}(r)\equiv V_{b}(r)$, we are able to
reproduce   the synthetic rotation curves $V_{syn}(r)$ at the level of their
r.m.s.   For $r>>6 R_d$, i.e. beyond the region described by the  
URC, the two velocity profiles progressively differ.

The values of  $r_0$ and $\rho_0$ from the URC   agree with the extrapolation
at high masses  of the scaling law $\rho \propto r_0^{-2/3} $ [3] 
established for objects with core radii $r_0$ ten times  smaller (see Fig.
4). Let us notice that the core radii are very   large: $r_0 \gg R_d$ so that
an   ever-rising halo RC cannot be excluded by the data. 
Moreover, the disk-mass vs. central halo density  relationship $\rho_0
\propto M_d^{-1/3}$, found for   dwarf galaxies [3],  according to which the
densest halos harbor the least massive disks,   holds also for disk systems of
stellar mass up to   $10^{11} {\rm M}_\odot$  (see Fig. 4).

\begin{figure}[h] 
\begin{center} 
\includegraphics[width=.87\textwidth, height=.7\textwidth]{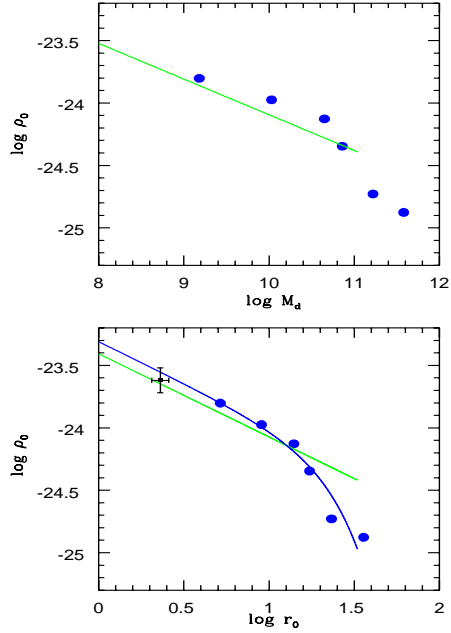}
\end{center} 
\caption[]{({\bf  up}) Disk mass (in solar units) {\it vs} central halo
density $\rho_0$  (in ${\rm g/cm}^3$) for normal spirals ({\it filled circles}). 
 The straight line is from [3]({\bf  bottom}) central density {\it vs} core radii (in
kpc) for normal  spirals ({\it filled circles}).                        
The   straight line and the point                        
 are   from the dwarfs sample of  [3].   The curved
 line is:  $\rho_0=5\times 10^{-24}r_0^{-2/3}\exp {-(r_0/27)^2} g/cm^3$.  }
\end{figure}

The above relationship shows a curvature at the highest masses/lowest
densities  that can be related to the existence of  
an upper limit in the dark halo mass $M_{200}$ 
\footnote{The virial  halo mass is given by $M_{200}\equiv 200 \times 4\pi/3
\rho_c R_{200}^3 \Omega_0 (1+z^3) g(z) $  with $z$ the formation redshift,
$R_{200} $ the virial radius, for $g(z)$   see e.g. [2]; the critical   density
is defined as: $\rho_c\equiv 3/(8 \pi) G^{-1} H_0^2$.} 
which is  evident by the sudden decline of the baryonic \emph{mass} function
of   disk galaxies at $M_{d}^{max}= 2\times 10^{11}M_\odot$ [26], that
implies a maximum halo mass of  $M_{200}^{max} \sim \Omega_0/\Omega_{b} 
M_{d}^{max}$,  where $\Omega_0$ and $\Omega_{b} \simeq 0.03$  [e.g. 5]  are
the matter and baryonic  densities of the Universe in units 
of critical density. From the definition of $M_{200}$, by means 
of eq. (6) and (8), we can write $M_{200}$ in terms of the  ``observable"
quantity $M_{0}$:  $M_{200} = \eta M_0$. 
For $(\Omega_0, z)=(0.3,3)$, $\eta \simeq 12$;  
notice that there is a mild dependence of $\eta$  
on $z$ and $\Omega_0$ which is irrelevant for the present study.  
From simple manipulation of previous equation- we obtain an upper limit for the central
density, $\rho_0 <1\times 10^{-20}  (r_0/{\rm kpc})^{-3} \ {\rm g /cm}^{3}$,
 which implies a lack of objects with $\rho_0> 4\times  10^{-25}\ {\rm
g/cm}^{3}$ and $r_0>30 \ {\rm kpc}$, as is evident in Fig. 4.   Turning the
argument around, the deficit of objects with  $M_d \sim M_d^{max} $ and
$\rho_0> 4\times 10^{-25}~{\rm g/cm}^{3}$, suggests   that, at this mass
scale,   the total-to-baryonic density ratio nears  the cosmological value
$\Omega/\Omega_{b}\simeq 10$.    

\subsection{Testing CDM}  
 
Out to two optical radii, the Burkert density profile reproduces,  
for the whole spiral luminosity sequence, the DM halos mass distribution. 
This density profile, though at very large radii coincides 
with the NFW profile, approaches a constant, finite density value at the center, in a way 
consistent with an isothermal distribution. This is in contradiction to  
CDM halo properties which predict [e.g. 10] that the
velocity  dispersion $\sigma$ of the 
dark matter particles decreases towards the center to reach  
$\sigma \rightarrow 0$ for $r \rightarrow 0$ . 
The dark halo inner regions,  
therefore, cannot be considered as kinematically cold structures  
but rather as  
``warm" regions with size $ r_0 \propto \rho_0^{-1.5}$.  
The halo core sizes are very large: $r_0 \sim 4-7 R_d$. 
Then, the boundary of the core region is well beyond the region  
where the stars are located and, 
as in [7], even at the outermost observed radius 
there is not the slightest evidence that dark halos converge 
to a $\rho \sim r^{-2}$ (or a steeper) regime.  

\section{Individual RC's and CDM}
 
To derive the halo density from an individual rotation curve is certainly complicated, 
however,  the belief according to which   RC's lead to ambiguous  halo  mass
modeling [e.g. 28] is incorrect. In fact this is 
true only for rotation curves of low  spatial resolution,
i.e. with  $< 3$ measures  per exponential disk length--scale $R_d$,
as for most of HI RC큦. Since the parameters of the
galaxy structure  are very sensitive to the  {\it
shape} of the rotation curve in the region $0<r<R_d$, 
 that corresponds to the region of the RC  steepest rise, then the
mass model cannot  be inferred if such a region  is poorly sampled and/or
radio beam--biased. Instead,   high--quality {\it optical} RC큦 with
tens of independent measurements in  the critical region 
probe the halo mass distribution  and resolve their structure.  
 Since the dark component can be better traced when the disk contributes 
to the dynamics in a modest way, it is convenient to investigate 
DM--dominated objects, like dwarf and low surface brightness (LSB) galaxies. 
It is well known that for the latter there are claims of dark 
matter distributions with regions of constant density well different from  
the cusped density distributions of the Cold Dark Matter 
scenario [e.g. 8, 13, 3, 4, 11, 12, 27].  However, these results are far from
certain being {\it 1)} under the (unlikely) {\it caveat} that  the low spatial
resolution of the RC큦 does not bias  the derived mass model and {\it 2)}
uncertain, due to the limited amount of  available kinematical data [see 29].
Since most of the properties of cosmological halos are claimed universal, 
we concentrate on a small and particular  sample of RC큦, that, nevertheless, 
reveal  the properties of the DM halos around spirals. A more useful strategy
has been to investigate  a number of high--quality {\it optical} rotation
curves  of {\it low luminosity } late--type spirals, with $I$--band absolute 
magnitudes $-21.4<M_I<-20.0$ and that $100 <V_{opt}< 170$ km s$^{-1}$. 
Objects in this  luminosity/velocity range are DM dominated [e.g. 20]  but
their RC's, measured at an angular resolution of $2^{\prime  \prime}$, have a
spatial resolution of $w\sim 100 (D/10 \ {\rm Mpc})$  pc and $n_{data}\sim
R_{opt}/w$ independent measurements.  For nearby galaxies: $ w<< R_{d}$ and
$n_{data}>25$.  Moreover, we select RC's of bulge--less systems, so that the 
stellar disk is the only baryonic component for $r \lsim R_d$. 

In detail, we take from [19] the rotation curves of 
the `excellent' subsample  of $80$ galaxies, which are suitable for an accurate
mass modeling. In fact,  these RC's properly trace the gravitational potential
in that: {\it 1)}  data extend at least to the optical radius, {\it 2)} they
are smooth and  symmetric, {\it 3)} they have small {\it rms}, {\it 4)} they
have high  spatial resolution and a homogeneous radial data coverage, i.e.
about  $30-100$ data points homogeneously distributed with radius and between
the  two arms. From this subsample we extract 9 rotation curves of low 
luminosity galaxies ($5 \times 10^9 L_{\odot} <L_I< 2 \times 
10^{10} L_{\odot}$; $100<V_{opt}< 170$ km s$^{-1}$), with their 
$I$--band surface luminosity being an (almost) perfect radial 
exponential. These two last criteria, not indispensable to 
perform the {\it mass} decomposition, are however required to 
infer the  dark halo {\it density}  distribution.  Each RC has $7-15$
velocity points inside  $R_{opt}$, each one being the average of $2-6$
independent data. The RC  spatial resolution is better than $1/20\ R_{opt}$,
the velocity r.m.s.  is about $3\%$ and the RC's logarithmic derivative is
generally known within  about 0.05.

\subsection{ Halo Density Profiles}
  
We model the mass distribution as the sum of two components: a stellar disk 
and a spherical dark halo. By assuming centrifugal equilibrium under the 
action of the gravitational potential, the observed circular velocity can be 
split into these two components:  

\begin{equation}  
V^2(r)=V^2_D(r)+V^2_H(r)  
\end{equation}  
By selection, the objects are bulge--less and the stellar  
component is distributed like an exponential thin disk. Light traces the  
mass via an assumed radially constant mass--to--light ratio.  
In the r.h.s of (9) we neglect the gas contribution $V_{gas}(r)$ since  
in normal spirals it is usually modest within the optical region [21,  Fig.
4.13]: $ \beta_{gas}\equiv (V^2_{\rm gas}/V^2)_{R_{opt}} \sim 0.1$.  
Furthermore, high resolution HI observations show that in  low luminosity
spirals:  $V_{gas}(r) \simeq 0$ for $r < R_d$ and  $V_{gas}(r) \simeq (20
\pm 5) (r-R_d) / 2 R_d $ for   $R_d \leq r \leq 3R_d $.  Thus, in the optical
region: {\it i)} $V_{gas}^2(r)<<V^2(r)$ and {\it ii)} 
$d[V^2(r)-V^2_{gas}(r)]/dr \gsim 0$. This last condition implies that by  
including $V_{gas}$ in the r.h.s. of (9) the halo velocity profiles  would
result {\it steeper} and then the core radius in the halo density   {\it
larger}. Incidentally, this is not the case for dwarfs and LSB큦: most  of
their kinematics is affected by the HI disk gravitational pull in such  a way
that neglecting it could bias the determination of the DM density.    
The circular velocity profile of the disk is given by (3) and the DM halo will
have the form given by (4). Since we normalize (at  $R_{opt}$) the velocity
 model $(V_h^2+V^2_d)^{1/2}$ to the observed rotation 
speed $V_{opt}$, $\beta$ enters explicitly in the halo velocity model  
and this reduces the free parameters of the mass model to two. 
 
It is important to remark that, out to $R_{opt}$, the proposed Constant 
Density Region (CDR) mass  model of (4) is instead  {\it neutral}  with respect to   
all the  proposed  models. Indeed, by varying $\beta$ and $a$, we  
can  efficiently reproduce the maximum--disk, the solid--body, the 
no--halo, the all--halo, the CDM and the core-less--halo models. For 
instance, CDM halos with concentration parameter $c=5$ and $r_s=R_{opt}$  are
well fit by (4) with $a \simeq 0.33$.   

For each galaxy, we determine the values of the parameters $\beta $ and $a$ 
by means of a $\chi ^2$--minimization fit to the observed rotation curves: 
 \begin{equation} 
V^2_{model}(r; \beta , a) = V^2_d (r; \beta) + V^2_d (r; \beta , a) 
\end{equation} 
A central role in discriminating among the different mass decompositions is  
played by the derivative of the velocity field $dV/dr$. It has been shown  
[e.g. 18] that by taking into   account the logarithmic gradient of the
circular velocity field defined as:  $\nabla (r)\equiv \frac{d \log V(r)}{d
\log r}  $  one can retrieve the  crucial  information  stored in the shape of
the rotation curve.   Then,  we set the    $\chi^2$-s  as the sum of those
evaluated on velocities and  on logarithmic  gradients: 
$ \chi^2_V =\sum^{n_V}_{i=1}\frac{V_i-V_{model}(r_i; \beta,a)}  {\delta V_i}$ 
and  $\chi^2_{\nabla} = \sum^{n_{\nabla}}_{i=1}\frac{\nabla(  
r_i)-\nabla_{model}(r_i; \beta,a)}{\delta \nabla_i}$,
with  $\nabla_{model} (r_i, \beta,a)$ given  from the above equations.
The parameters  of the mass models are finally  obtained   by minimizing the
quantity $\chi^2_{tot} \equiv \chi^2_V+ \chi^2_{\nabla}$.

\begin{figure}[h] 
\begin{center} 
\includegraphics[width=.4\textwidth]{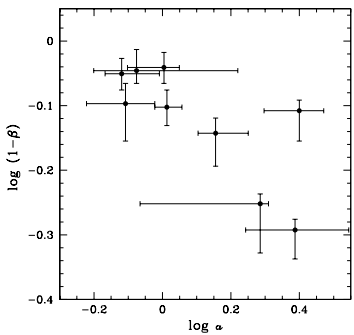} 
\end{center} 
\vspace {-0.2truecm}
\caption[]{Halo parameters ($a$ is in units  of $R_{opt}$) with their uncertainties} 
\end{figure}

The parameters of the best--fit models   are shown in Fig. 5. They are very
well specified: the allowed values   span a small and continuous region of the
($a$, $\beta$)   space. We get a ``lowest" and a ``highest" halo velocity  
curve by subtracting from $V(r)$ the maximum and the minimum disk  
contributions $V_d(r)$ obtained by substituting in (3) the parameter 
$\beta$ with $\beta_{best}+\delta\beta$ and $\beta_{best}-\delta\beta$, 
respectively. 
The derived mass models are shown in Fig. 6, alongside with the separate  
disk and halo contributions. It is then obvious that the halo curve is  
steadily increasing, almost linearly, out to the last data point.  
The disk--contribution $\beta $ and the halo core radius $a$ span a range  
from 0.1 to 0.5 and from 0.8 to 2.5, respectively. In each object the  
uniqueness of the resulting halo velocity model can be realized by the  
fact that the maximum--disk and minimum--disk models almost coincide.  
Remarkably, we find that the size of the halo density core is  
always greater than the disk characteristic scale--length  
$R_d$ and it can extend beyond the disk edge (and the region  
investigated).

\begin{figure}[h] 
\vspace{-1.6truecm}
\begin{center}
\includegraphics[width=12.5truecm,height=17truecm]{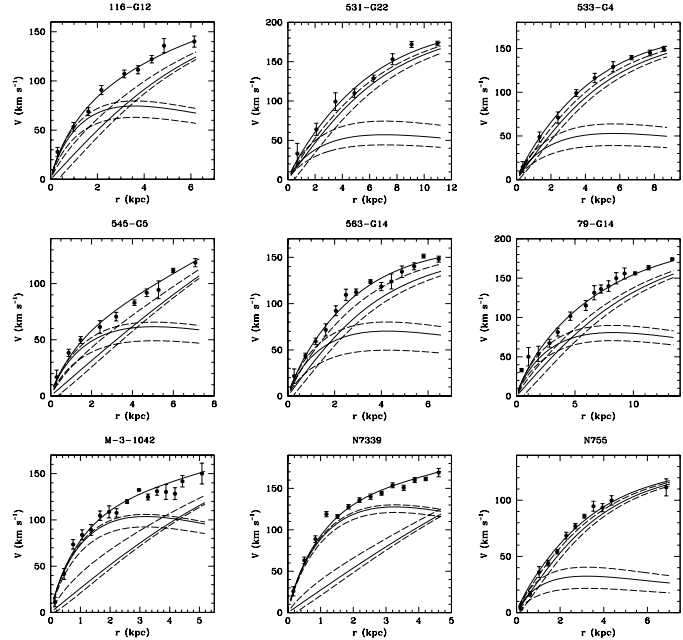} 
\end{center}
\vspace{-6.7truecm}

\caption[]{CDR model fits ({\it thick solid line}) to the RC큦
({\it points with errorbars}). Thin solid lines represent the   disk and halo
contributions. The maximum disk and the minimum   disk solutions are also
plotted ({\it dashed lines})}  
\end{figure}

\subsection{Testing CDM} 
 
In Fig. 7  we show the halo velocity profiles 
for the nine galaxies. The halo circular velocities are normalized to their 
values at $R_{opt}$ and expressed as a function of the normalized radius 
$r/R_{opt}$. These normalizations allow a meaningful comparison between 
halos of different masses. It is then evident that the halo circular velocity, in every 
galaxy, rises almost linearly with radius, at least out to the disk edge: 
$V_h(r) \propto r$ for  $0.05R_{opt} \lsim r \lsim R_{opt}$. 

The halo density profile has a well defined (core) radius within which 
the density is approximately constant. This is inconsistent with the 
singular halo density distribution emerging in the Cold Dark Matter (CDM) 
 halo formation scenario. More precisely, since the CDM 
halos are, at small radii, likely more cuspy than the NFW profile: 
$\rho_{CDM}\propto r^{-1.5}$ [e.g. 14], the steepest CDM halo 
velocity profile $V_h(r) \propto r^{1/4}$ results too shallow with 
respect to observations.  
Although the mass models of (4) converge to a distribution  with an inner
core rather than with a central spike, 
it is worth, given the importance of such result,  also checking in a direct
way the (in)compatibility of the CDM  models with galaxy  kinematics.  We
assume the NFW two--parameters functional form for the halo density  [15, 16,
17], given by (1). 
Though N--body simulations and semi-analytic investigations indicate  that
the two parameters $c_{\rm vir}$ and $r_s$ correlate,  they are left 
independent to increase the chance of a good  fit. For the object under study 
a generous   halo mass $M_{vir}$  upper limit is 
$M_{up}= 2 \times 10^{12} M_\odot$.

\begin{figure}[h] 
\vspace{-0.7truecm}
\begin{center} 
\includegraphics[width=.5\textwidth]{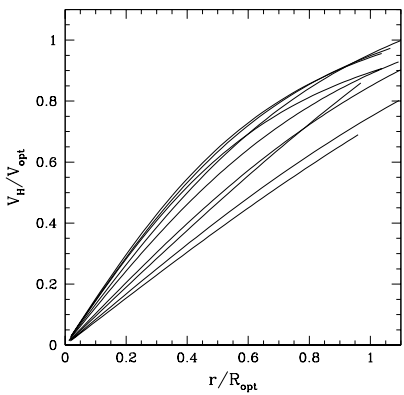} 
\end{center} 
\vspace{-0.5truecm}
\caption[]{The halo velocity profiles of the sample spirals.  
$V_h(r)$ rises almost linearly with radius: the DM halo density 
remains  approximately constant} 
\end{figure} 

\noindent The fits to the data are shown in  Fig. 8 and  compared with  the NFW models: 
for seven out of  nine  objects the latter are unacceptably worse than the
CDR solutions, moreover  in all objects,   the CDM  virial mass
 is  too high  high: $M_{vir}\sim 2 \times 10^{12} M_{\odot}$ and  the resulting
  disk  mass--to--light ratio too low.
 The inadequacy of the CDM model for our sample galaxies is even more 
evident if one performs the fit after removing the  constraint on virial
mass. In fact, good fits are obtained only for very low values of the
concentration  parameter ($c_{\rm vir} \simeq 2$) and for ridiculously  large
virial velocities  and masses ($V_{\rm vir} \simeq 600-800 \ {\rm km \
s^{-1}}; M_{\rm vir}  \simeq 10^{13}-10^{14} {\rm M}_{\odot}$). These results
can be explained  as effect of the attempt, by  the minimization routine,  to
fit  the NFW velocity profile  ($V(r)  \propto r^{0.5}$)  to data
intrinsically linear in $r$.   
 
\section{Conclusions: an Intriguing Evidence} 
 
The dark halos around spirals emerge as an one--parameter family;  
it is relevant that the order parameter (either the central density or the core
radius)  correlates with the luminous mass. However, we do not know how it is 
related to the global structural properties of the dark halo, like the 
virial radius or the virial mass. The halo RC, out to $6R_d$, is
completely determined by parameters,   i.e. the central core density 
and the core radius, which are not defined in present gravitational
instability/ hierarchical clustering  scenario.
In fact the location of spiral galaxies in the parameter space of virial 
mass, halo central density and baryonic mass,  determined by  
different processes on different scales ,  degenerates with no doubt into a
single curve (see Fig. 4), we recall that:   $\rho_0 = {\pi\over {24}} ~
M_{200}/r_0^{3}$ and   $M_d=G^{-1}\beta V^2_{opt}R_{opt}$,  of  
difficult interpretation within the standard theory of galaxy formation. 
 \begin{figure}[h]
\vspace{-2truecm} 
\begin{center}
\includegraphics[width=12.5truecm, height=19truecm]{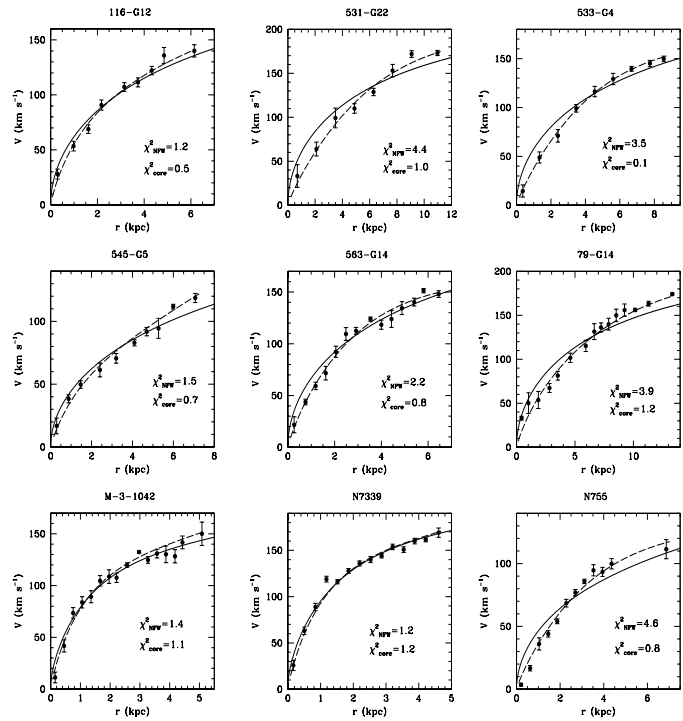} 
\end{center}
\vspace{-7.5truecm} 
\caption[]{NFW best--fits {\it solid lines} of the rotation curves  {\it
(filled circles)} compared with   the CDR fits {\it (dashed lines)}. The  
$\chi^2$ values are also indicated} 
\end{figure} 
Crucial  insight  has come from  disk--halo density  decompositions   of 
a number of  disk galaxies. 
These galaxies have a relevant amount of dark matter: the 
contribution of the luminous matter to the dynamics is small and 
it can be easily  taken into account. Moreover, the high spatial resolution 
of the available rotation curves allows us to derive  the separate dark 
and luminous density profiles. 
We find that dark matter halos have a constant central 
density region whose size exceeds the stellar disk length--scale $R_d$. 
As result,  the halo profiles disagree with the cuspy density 
distributions typical of CDM halos which, therefore, fail to account for the actual DM 
velocity data.  

Pointing out  that   a  review  on the  various  efforts aimed to cope   with
the core radii evidence will be published elsewhere,   we conclude by stressing that, 
for {\it any}  theory of galaxy formation,  time is come 
to seriously consider that stellar  disks (and perhaps also stellar spheroids) 
lay down in dark  halos of constant density.

\end{document}